\documentclass[prl,twocolumn,showpacs,superscriptaddress]{revtex4}

\usepackage{amsmath,amssymb}
\usepackage{verbatim}
\usepackage{graphicx}
\usepackage{hyperref}
\usepackage{color}

\DeclareFontFamily{OT1}{rsfs}{}
\DeclareFontShape{OT1}{rsfs}{m}{n}{ <-7> rsfs5 <7-10> rsfs7 <10->rsfs10}{} 
\DeclareMathAlphabet{\mycal}{OT1}{rsfs}{m}{n}

\newcommand{\be}[1]{ \begin{equation}\label{#1} }
\newcommand{\ee}{\end{equation}}
\newcommand{\bea}[1]{\begin{eqnarray}\label{#1} }
\newcommand{\eea}{\end{eqnarray}}
\newcommand{\eq}[2]{\begin{equation} #1 \label{#2} \end{equation}}
\newcommand{\eps}{\varepsilon}

\newcommand{\ga}{\gamma}

\newcommand{\la}{\lambda}
\newcommand{\si}{\sigma}

\newcommand{\Ga}{\Gamma}

\newcommand{\Om}{\Omega}

\newcommand{\La}{\Lambda}

\DeclareMathOperator{\extdm}{d}
\newcommand{\extd}{\extdm \!}

\begin{document}

\title{Cosmic evolution from phase transition of 3-dimensional flat space}

\author{Arjun Bagchi}
\email{arjun.bagchi@ed.ac.uk}
\affiliation{School of Mathematics, University of Edinburgh, Edinburgh EH9 3JZ, United Kingdom}
\affiliation{Center for Theoretical Physics, Massachusetts Institute of Technology, Cambridge, MA 02139, USA}
\author{Stephane Detournay}
\email{sdetourn@physics.harvard.edu}
\affiliation{Center for the Fundamental Laws of Nature, Harvard University, Cambridge, MA 02138, USA}

\author{Daniel Grumiller}
\email{grumil@hep.itp.tuwien.ac.at}
\affiliation{Institute for Theoretical Physics, Vienna University of Technology, Wiedner Hauptstrasse 8--10/136, A-1040 Vienna, Austria}

\author{Joan Sim\'on}
\email{j.simon@ed.ac.uk}
\affiliation{School of Mathematics, University of Edinburgh, Edinburgh EH9 3JZ, United Kingdom}

\date{\today}

\preprint{TUW--13--06}

\begin{abstract} 
Flat space cosmology spacetimes are exact time-dependent solutions of 3-dimensional gravity theories, such as Einstein gravity or topologically massive gravity. We exhibit a novel kind of phase transition between these cosmological spacetimes and the Minkowski vacuum. At sufficiently high temperature (rotating) hot flat space tunnels into a universe described by flat space cosmology.
\end{abstract}


\maketitle

\paragraph{Introduction.} Phase transitions are ubiquitous in physics, with numerous applications in condensed matter physics, particle physics and cosmology. Interestingly, phase transitions can occur even between different spacetimes, for instance between black hole spacetimes and (hot) empty space \cite{Hawking:1982dh}. In this work we exhibit a novel type of phase transition between cosmological spacetimes and (hot) flat space in three spacetime dimensions. 

The existence of a phase transition is quite surprising given that flat space in three dimensions ($\varphi\sim\varphi+2\pi$)
\eq{
\extd s^2 = -\extd t^2 + \extd r^2 + r^2\,\extd\varphi^2
}{eq:phase} 
has few interesting features at first glance. It arises as an exact solution of the vacuum Einstein equations $R_{\mu\nu}=0$.
The Euclidean signature version of flat space allows to introduce a finite temperature by periodically identifying the Euclidean time (possibly with a rotation in $\varphi$). We call this hot flat space (HFS). 

The other spacetime we shall be concerned with is flat space cosmology (FSC).
FSC spacetimes \cite{Cornalba:2002fi,Cornalba:2003kd} [$\La(\tau):=1+(E\tau)^2$, $y\sim y + 2\pi r_0$] 
\eq{
\extd s^2=-\extd\tau^2 + \frac{(E\tau)^2\,\extd x^2}{\La(\tau)} + \La(\tau)\,\Big(\extd y+\frac{(E\tau)^2}{\La(\tau)}\,\extd x\Big)^2
}{eq:phase2}
are locally flat \cite{Deser:1983tn} time-dependent exact solutions of the vacuum Einstein equations \cite{Barnich:2012xq,Bagchi:2012xr}. For positive (negative) $\tau$ they describe expanding (contracting) universes from (towards) a cosmological horizon at $\tau=0$. The parameter $E$ has inverse length dimension and corresponds physically to the temperature associated with FSC \cite{Cornalba:2002nv}.

The main purpose of the present work is to exhibit a phase transition between the (Euclidean versions of the) spacetimes \eqref{eq:phase} and \eqref{eq:phase2} within Einstein gravity and more general gravitational theories in three dimensions. Thus, remarkably time-dependent cosmological spacetimes can emerge from flat space by heating up the latter. 

\paragraph{Flat space cosmological spacetimes.}
FSC spacetimes \eqref{eq:phase2} are shifted-boost orbifolds of $\mathbb{R}^{1,\,2}$ \cite{Cornalba:2002fi,Cornalba:2003kd} and correspond to flat space analogues of non-extremal rotating Ba\~nados--Teitelboim--Zanelli (BTZ) black holes \cite{Banados:1992wn} in Anti-de~Sitter (AdS).
In flat space chiral gravity \cite{Bagchi:2012yk} FSC spacetimes are conjectured to be dual to non-perturbative states, again in full analogy to the role played by BTZ black holes in AdS quantum gravity. 
Their Bekenstein-Hawking entropy can be matched by a formula counting the asymptotic growth of states in the putative dual %Galilean conformal 
field theory \cite{Bagchi:2012xr,Barnich:2012xq}. 
%They also allow a microscopic derivation of their entropy by counting of states in the dual Galilean conformal field theory \cite{Bagchi:2012xr,Barnich:2012xq}. 
%
It is useful for our purposes to represent FSC \eqref{eq:phase2} in terms of different coordinates. We make the coordinate transformation $\hat r_+ t = x$, $r_0\varphi=y+x$ and $(r/r_0)^2=1+(E\tau)^2$ with $E=\hat r_+/r_0$ and $\varphi\sim\varphi+2\pi$.
\eq{
\extd s^2 = \hat r_+^2\,\extd t^2 - \frac{r^2\,\extd r^2}{\hat r_+^2\,(r^2-r_0^2)} + r^2\,\extd\varphi^2 - 2\hat r_+ r_0\,\extd t \extd\varphi
}{eq:phase1}
With no loss of generality we assume $r_0, \hat r_+ >0$. These solutions are compatible with asymptotically flat boundary conditions \cite{Barnich:2006av,Bagchi:2012yk}. In the absence of sources, \eqref{eq:phase1} is the most general zero mode solution of the vacuum Einstein equations \cite{Barnich:2012aw}. At vanishing $r$ closed null curves are encountered, so that the locus $r=0$ corresponds to a singularity in the causal structure. 

This singularity is screened by a cosmological horizon at the surface $r=r_0$, so that the region $r\geq r_0$ is regular. The horizon's surface gravity determines its Hawking-temperature $T=\beta^{-1}$ as 
\eq{
T = \frac{\hat r_+^2}{2\pi r_0}\,.
}{eq:phase5}
The angular velocity $\Om$ of the horizon is given by 
\eq{
\Om = \frac{\hat r_+}{r_0}\,.
}{eq:phase5a}
The result for Hawking-temperature \eqref{eq:phase5} agrees with the corresponding one by Cornalba, Costa and Kounnas \cite{Cornalba:2002nv}, who calculated thermal radiation from cosmological particle production in the time-dependent background \eqref{eq:phase2}.

\paragraph{Strategy of the calculation.}
Given some values of temperature \eqref{eq:phase5} and angular velocity \eqref{eq:phase5a} we pose the question which of the spacetimes \eqref{eq:phase}, \eqref{eq:phase1} is preferred thermodynamically. To this end we continue to Euclidean signature and compare which of the smooth Euclidean solutions has smaller free energy. 

Free energy can then be derived from the canonical partition function
\eq{
Z(T,\,\Omega) = \int{\cal D}g\,e^{-\Gamma[g]} = \sum_{g_c} e^{-\Gamma[g_c(T,\,\Omega)]}\,\times Z_{\rm fluct}
}{eq:phase6}
where the path integral is performed over all continuous Euclidean metrics $g$ compatible with the boundary conditions enforced by the temperature $T$ and angular velocity $\Omega$. In the semi-classical approximation, the leading contribution comes from the Euclidean action $\Gamma$ evaluated on smooth classical solutions $g_c$ compatible with the boundary conditions. We are not concerned here with subleading contributions from fluctuations encoded in $Z_{\rm fluct}$.

\paragraph{Smooth Euclidean saddle points.}
The Euclidean version of flat space \eqref{eq:phase} is simple, but we need also the Euclidean continuation of FSC \eqref{eq:phase1}. A natural choice is
\eq{
t = i \tau_{\textrm{\tiny E}}  \qquad \hat r_+ = - i r_+
}{eq:angelinajolie}
which then leads to Euclidean FSC
\eq{
\extd s^2_{\textrm{\tiny E}} = r_+^2\big(1-\tfrac{r_0^2}{r^2}\big)\extd\tau_{\textrm{\tiny E}} ^2 + \frac{\extd r^2}{r_+^2(1-\tfrac{r_0^2}{r^2})} + r^2\big(\extd\varphi - \frac{r_+ r_0}{r^2}\extd\tau_{\textrm{\tiny E}}\big)^2\,.
}{eq:phase24}
Requiring the absence of conical singularities on the FSC horizon fixes the periodicities of the angular coordinate $\varphi$ and Euclidean time $\tau_{\textrm{\tiny E}}$.
Consider $r^2=r_0^2 + \epsilon\rho^2$. In the near-horizon $(\epsilon\to 0)$ approximation, the Euclidean metric \eqref{eq:phase24} can be written as
\begin{equation}
  \extd s_{\textrm{\tiny E}}^2 \approx \frac{\epsilon}{r_+^2}\,\Big(\extd\rho^2 + \rho^2\ \frac{r_+^4}{r_0^2}\ \extd\tau_{\textrm{\tiny E}}^2\Big) + r_0^2\,\Big(\extd\varphi-\frac{r_+}{r_0}\extd\tau_{\textrm{\tiny E}}\Big)^2\,.
\label{eq:phase7}
\end{equation}
Smoothness requires the identifications
\begin{equation}\label{Per}
  \tau_{\textrm{\tiny E}} \sim \tau_{\textrm{\tiny E}} + \frac{2\pi r_0}{r_+^2} = \tau_{\textrm{\tiny E}} + \beta \quad \varphi \sim \varphi + \frac{2\pi}{r_+} = \varphi + \beta \Omega\,,
\end{equation}
since the term proportional to $\epsilon$ demands a precise $\tau_{\textrm{\tiny E}}$ periodicity and the transverse direction $\varphi - \tau_{\textrm{\tiny E}} r_+/r_0$ must stay fixed as one moves around the thermal $\tau_{\textrm{\tiny E}}$ circle. The expressions for Hawking-temperature $T=\beta^{-1}=r_+^2/(2\pi r_0)$ and angular velocity $\Omega=r_+/r_0$ agree with their Minkowski counterparts \eqref{eq:phase5}, \eqref{eq:phase5a} as well as with the flat limit expressions of the ones for inner horizon BTZ thermodynamics \cite{Castro:2012av,Detournay:2012ug}.

\paragraph{Defining the ensemble.} 
We declare two Euclidean saddle points to be in the same ensemble if
\begin{enumerate}
 \item they have the same temperature $T=\beta^{-1}$ and angular velocity $\Om$ given by \eqref{eq:phase5} and \eqref{eq:phase5a}, respectively,
 \item the two metrics obey flat space boundary conditions \cite{Barnich:2006av,Bagchi:2012yk}, and
 \item the solutions do not have conical singularities.
\end{enumerate}
Note that requirement 2.~is somewhat different from what we would usually assume, namely that the metrics asymptote to the same one at infinity. The peculiarities of the boundary conditions \cite{Barnich:2006av,Bagchi:2012yk} for flat space solutions imply that there are leading terms in the metric that can fluctuate like the $g_{tt}$ term.   Note finally that the absence of conical singularities does not automatically imply the absence of asymptotic conical defects! A crucial counterexample is FSC \eqref{eq:phase24}, which has an asymptotic conical defect if $r_+^2\neq 1$, since in the large $r$ limit $\extd s^2_{\textrm{\tiny E}} = r_+^2\extd\tau_{\textrm{\tiny E}} ^2 + \extd r^2/r_+^2 + r^2\extd\varphi^2 + \dots$, where $\varphi$ is $2\pi$-periodic \footnote{%
This property can be understood as limiting case of geometric properties between two BTZ horizons.
The Euclidean geometry between the inner and the outer horizon in general has a conical singularity on either of the two horizons, depending on how the periodicities are fixed. Suppose that we fix them such that the inner horizon is free from a conical singularity. Then the outer horizon has a conical defect and a conical singularity. In the flat space limit the outer horizon is pushed toward infinity, so that the conical singularity is not part of the manifold. Nevertheless, asymptotically there is a conical defect.
}. On the other hand, Euclidean HFS, 
\be{eq:phase36}
ds^2_{\textrm{\tiny HFS}} = \extd\tau^2_{\textrm{\tiny E}} + \extd r^2 + r^2 \extd\varphi^2
\ee
has no conical defects since it has periodicities 
$(\tau_{\textrm{\tiny E}},\, \varphi) \sim (\tau_{\textrm{\tiny E}},\, \varphi + 2 \pi) \sim (\tau_{\textrm{\tiny E}} + \beta,\, \varphi + \Phi)$
where inverse temperature $\beta=T^{-1}$ and angular potential $\Phi=\beta\Omega$ are given by \eqref{eq:phase5}. 

\paragraph{Cosmic phase transition in Einstein gravity.} The considerations above are valid for any 3-dimensional (3d) gravity theory supporting flat space boundary conditions. From now on we focus on the simplest such theory, namely Einstein gravity. Its Euclidean action reads \footnote{%
We use the standard Einstein--Hilbert action with one half of the Gibbons--Hawking--York boundary term \cite{Gibbons:1976ue,York:1972sj} for calculating the on-shell action. 
It turns out \cite{Scholler:2013} that flat space boundary conditions \cite{Barnich:2006av,Bagchi:2012yk} require such a boundary term for a well-defined variational principle.
A detailed analysis of the variational principle is in preparation with Friedrich Sch\"oller.
}
\eq{
\Ga = -\frac{1}{16\pi G}\,\int\!\extd^3x\sqrt{g}\,R - \frac{1}{16\pi G}\,\lim_{r\to\infty}\,\int\!\extd^2x\sqrt{\ga}\,K\,.
}{eq:phase9}
Here $G$ is the Newton constant, $\ga$ the determinant of the induced metric at the asymptotic boundary $r\to\infty$ and $K$ the trace of extrinsic curvature.

On-shell the bulk term vanishes in \eqref{eq:phase9}.
HFS \eqref{eq:phase36} yields $\sqrt{\ga}=r$ and $K=1/r$.
FSC \eqref{eq:phase7} yields $\sqrt{\ga}=r_+r+{\cal O}(1/r)$ and $K=r_+/r+{\cal O}(1/r^3)$.
Thus, we obtain on-shell
\eq{
\Gamma_{\textrm{\tiny HFS}} = -\frac{\beta}{8G}\qquad \Gamma_{\textrm{\tiny FSC}} = -\frac{\beta\, r_+^2}{8G}=-\frac{\pi r_0}{4G}\,.
}{eq:phase10}
Plugging the on-shell actions \eqref{eq:phase10} into \eqref{eq:phase6} establishes the respective canonical partition functions $Z(T,\,\Omega)$.
The free energy is obtained from $F(T,\,\Omega)=-T\,\ln Z$, where $T=r_+^2/(2\pi r_0)$ is the Hawking-temperature.
\eq{
F_{\textrm{\tiny HFS}} = T\,\Gamma_{\textrm{\tiny HFS}} = -\frac{1}{8 G} 
\qquad F_{\textrm{\tiny FSC}} = T\,\Gamma_{\textrm{\tiny FSC}}  = - \frac{r_+^2}{8 G} 
}{eq:phase35}
So our main conclusion is that there is a phase transition between HFS and FSC as summarized below ($r_+>0$):
\bea{eq:phase39}
r_+ >1: && {\mbox{FSC is the dominant saddle.}} \nonumber \\
r_+ <1: && {\mbox{HFS  is the dominant saddle.}} \\
r_+ =1: && {\mbox{FSC and HFS coexist.}} \nonumber
\eea
The phase transition occurs at the critical temperature
\eq{
T_c = \frac{1}{2\pi r_0}\,= \frac{\Omega}{2\pi}\,.
}{eq:Tcrit}
%Thus, at sufficiently high temperature (rotation), HFS (FSC) `melts' (`falls apart') and tunnels into FSC (HFS).
Thus, at sufficiently high temperatures, HFS ``melts" and tunnels into FSC. Conversely, increasing rotation leads to higher critical temperatures making HFS more stable. So, at sufficiently high angular velocity, FSC ``falls apart" and tunnels to HFS.

\paragraph{Entropy.}
Given the free energy \eqref{eq:phase35} we can derive all thermodynamical variables of interest by standard methods. For HFS the free energy is constant, and thus all quantities besides temperature and angular rotation are trivial. In particular the entropy of HFS vanishes. By contrast, FSC has non-trivial thermodynamics \cite{Barnich:2012xq,Bagchi:2012xr}. As a consistency check on the validity of our result \eqref{eq:phase35} we show now that we recover the correct entropy.

Let us first rewrite the free energy in terms of temperature and angular rotation.
\eq{
F_{\textrm{\tiny FSC}}(T,\,\Omega) = -\frac{\pi^2\,T^2}{2G\,\Omega^2}
}{eq:phase13}
The thermodynamical entropy
\eq{
S = -\frac{\partial F_{\textrm{\tiny FSC}}}{\partial T}\Big|_{\Omega=\textrm{\tiny const.}} = \frac{2\pi r_0}{4G}
}{eq:phase14}
then coincides precisely with the Bekenstein--Hawking entropy, which in turn coincides with the entropy derived from a Cardy-like formula valid for Galilean conformal algebras \cite{Barnich:2012xq,Bagchi:2012xr}. Our result \eqref{eq:phase14} not only confirms the analysis of \cite{Barnich:2012xq,Bagchi:2012xr}, it strengthens their conclusions since we have derived above the Bekenstein--Hawking law from first principles, rather than assuming its validity.

Specific heat $C=T\partial S/\partial T=S=\pi^2T/(G\Om^2)$ is positive, which implies that the Gaussian fluctuations contained in $Z_{\rm fluct}$ in \eqref{eq:phase6} do not destabilize the system. Note that specific heat vanishes linearly with temperature as $T$ tends to zero, just like a free Fermi gas at low temperature.

\paragraph{First law.}
For FSC another thermodynamical quantity of interest is the angular momentum
\eq{
J = -\frac{\partial F_{\textrm{\tiny FSC}}}{\partial \Omega}\Big|_{T=\textrm{\tiny const.}} = - \frac{r_+ r_0}{4G}
}{eq:J}
which enters in the first law of thermodynamics
\eq{
\extd F = -S \,\extd T  - J \, \extd\Omega \,.
}{eq:firstlaw}
Integrating the first law yields
%\eq{
$F=U - TS - \Omega J = U$
%}{eq:integratedfirstlaw}
with the (non-positive) internal energy $U=\Omega J/2=-M$, where 
\eq{
M=\frac{r_+^2}{8G}
}{eq:mass}
is the mass parameter.
The first law is also obeyed by internal energy, $\extd U=T\,\extd S+\Omega\,\extd J$.
The unusual signs appearing here are reminiscent of inner horizon thermodynamics \cite{Larsen:1997ge, Cvetic:1997uw, Curir:1981uc, Castro:2012av, Detournay:2012ug}.

\paragraph{Matching the solutions via $S$-transformation.} 
We connect now FSC and HFS by the flat space analogue of a modular $S$-transformation in a Conformal Field Theory (CFT). This is useful for a field-theoretic interpretation of our results. 
The flat space $S$-transformation reads \footnote{%
The asymptotic symmetry algebra in 3d flat space is the infinite dimensional BMS$_3$ \cite{Bondi:1962,Sachs:1962,Barnich:2006av}. The field theory dual to 3d flat space has the same symmetry algebra, which is isomorphic to the 2-dimensional (2d) Galilean Conformal Algebra \cite{Bagchi:2010zz, Bagchi:2012cy} studied earlier in the context of non-relativistic AdS/CFT \cite{Bagchi:2009my, Bagchi:2009pe}. The microscopic derivation of the Cardy-like entropy formula in the field theory makes use of the counterpart of modular invariance in 2d CFTs. The latter can be derived either by looking at the limit of the 2d CFT modular transformation \cite{Bagchi:2012xr, Hotta:2010qi}, or by using the bulk symmetries directly (\cite{Barnich:2012xq}, following methods outlined in \cite{Detournay:2012pc}).
}
\be{eq:phase25}
S: \big(\beta,\, \Phi\big) \to \big(\beta',\, \Phi'\big) = \big(\frac{4 \pi^2 \beta}{\Phi^2},\, -\frac{4 \pi^2}{\Phi}\big)\,.
\ee
We start with the FSC metric \eqref{eq:phase24}. Changing coordinates
%\be{eq:phase26}
$r^2 = r_0^2 + r_+^2 r'^2$, $\tau_{\textrm{\tiny E}} = \tau'_{\textrm{\tiny E}}/r_+ - \varphi' r_0/r_+^2$, $\varphi = \varphi'/r_+$
%\ee
yields flat space
  $\extd s^2 = \extd\tau_{\textrm{\tiny E}}'^2 + \extd r'^2  + r'^2\,\extd\varphi'^2$.
In terms of the new coordinates, the periodicities read
 $(\tau_{\textrm{\tiny E}}',\varphi') \sim (\tau_{\textrm{\tiny E}}' - \beta',\, \varphi' + \Phi') \sim (\tau_{\textrm{\tiny E}}',\, \varphi' + 2 \pi)$
with
 $\beta' = 2 \pi r_0=4 \pi^2 \beta/\Phi^2$ and $\Phi' = 2 \pi r_+ = -4 \pi^2/\Phi$.
These are precisely the values obtained from the $S$-transformation \eqref{eq:phase25}.
Therefore, FSC with periodicities $(\beta, \,\Phi)$ is equivalent to HFS with $S$-dual periodicities $(\beta',\, \Phi')$. This is the flat space analogue of the AdS$_3$/CFT$_2$ statement that thermal AdS$_3$ with modular parameter $\tau$ is equivalent to a BTZ black hole with $S$-dual modular parameter $-1/\tau$ (see e.g.~\cite{Kraus:2006wn}).

\newcommand{\len}{L}

\paragraph{Consistency check.}
The analysis above lets us resolve a seemingly puzzling conceptual issue. In flat space \eqref{eq:phase} there appears to be no preferred scale, so how is it possible that there is a critical temperature? The key observation is that we are considering flat space with fixed angular rotation, which does provide a length scale $\len=2\pi r_0$. The critical temperature \eqref{eq:Tcrit} is reached precisely when the periodicity in Euclidean time is one in units of $\len$. We can interpret this property from a field theory perspective, where %the length scale 
$\len$ is associated with the twist of one of the cycles of the torus on which the field theory lives. 
Consistently, the critical temperature \eqref{eq:Tcrit} coincides with the self-dual point of the $S$-transformation \eqref{eq:phase25}.

\paragraph{Beyond Einstein gravity}
We generalize now our results to another interesting 3d theory of gravity, namely topologically massive gravity (TMG) \cite{Deser:1982vy}. 
\eq{
\Ga^{\textrm{\tiny TMG}} = \Ga - \frac{1}{32\pi G\,\mu}\,\int\extd^3x\,\textrm{CS}
}{eq:phase19}
If the Chern--Simons coupling constant $\mu$ tends to infinity we recover the Einstein gravity action \eqref{eq:phase9}.
The Chern--Simons term expressed in terms of the Christoffel symbols reads CS$\,=\epsilon^{\la\mu\nu}\,\Ga^\si_{\la\rho}\,\big(\partial_\mu\Ga^\rho_{\nu\si}+\tfrac23\,\Ga^\rho_{\mu\tau}\Ga^\tau_{\nu\si}\big)$.
TMG has all solutions of Einstein gravity, since any spacetime with vanishing Ricci tensor also has vanishing Cotton tensor, $C_{\mu\nu}=\eps_\mu{}^{\la\si}\nabla_\la\big(R_{\nu\si}-\frac14\,g_{\nu\si}R\big)=0$, and thereby trivially solves the field equations of TMG, $R_{\mu\nu} + \tfrac1\mu\,C_{\mu\nu}=0$. We assume with no loss of generality that $\mu$ is positive.

A complication in TMG is that it is not known how to compute free energy from the on-shell action. We proceed by assuming the validity of the first law of thermodynamics \eqref{eq:firstlaw} and then integrate it. For this we need the angular momentum at finite $\mu$, $J(\mu)=J-\tfrac1\mu M$, and entropy. The latter can be calculated using Solodukhin's conical deficit method \cite{Solodukhin:2005ah} or Tachikawa's generalization of the Wald entropy for theories with a gravitational Chern--Simons term 
\cite{Tachikawa:2006sz}.
\eq{
S^{\textrm{\tiny TMG}} = \frac{2\pi\,r_0}{4G} + \frac{1}{\mu}\,\frac{2\pi\,r_+}{4G}
}{eq:phase20}
The first term is compatible with the Einstein result \eqref{eq:phase14} obtained in the limit $\mu\to\infty$. The second term is compatible with the conformal Chern--Simons gravity (CSG) result obtained in the limit $G\to\infty$, $8\mu G = 1/k$, which we now derive. To this end, we exploit the flat space chiral gravity conjecture that the dual field theory is a chiral CFT with central charge $c=24k$ \cite{Bagchi:2012yk} and use a chiral version of the Cardy formula.
\eq{
S^{\textrm{\tiny CSG}} = 2\pi\,\sqrt{\frac{c\,h_L}{6}} = 4\pi k\, r_+
}{eq:phase21}
In the second equality we used the result for the Virasoro zero mode charge $h_L=k\,r_+^2$ \cite{Bagchi:2012yk}. 
Integrating the first law \eqref{eq:firstlaw} with the results above yields the free energy
\eq{
F^{\textrm{\tiny TMG}}_{\textrm{\tiny FSC}} = -\frac{\pi^2\, T^2}{2G\,\Om^2}\,\Big(1+\frac{\Om}{\mu}\Big)\,.
}{eq:phase18}

Comparing the free energies \eqref{eq:phase18} and $F^{\textrm{\tiny TMG}}_{\textrm{\tiny HFS}} = -\frac{1}{8G}$ we see that that there is again a phase transition between HFS and FSC as summarized below ($\mu,\, \Om,\, r_+>0$):
\begin{align}
r_+^2\big(1+ \tfrac{\Om}{\mu}\big) >1: & \;{\mbox{FSC is the dominant saddle.}} \nonumber \\
r_+^2\big(1+ \tfrac{\Om}{\mu}\big) <1: & \;{\mbox{HFS  is the dominant saddle.}} \label{eq:phase42}\\
r_+^2\big(1+ \tfrac{\Om}{\mu}\big) =1: & \;{\mbox{FSC and HFS coexist.}} \nonumber
\end{align}
Consequently, if $r_+$  is sufficiently large HFS is thermodynamically not the preferred spacetime and will tunnel to FSC. Thus, our phase transition is not a unique feature of Einstein gravity and arises also in TMG. The phase transition occurs at the critical temperature
\eq{
T_c^{\textrm{\tiny TMG}} = \frac{\Om}{2\pi}\,\frac{1}{\sqrt{1+\Om/\mu}}\,.
}{eq:phase43}

It could be interesting to extend our results to other 3d models, like New Massive Gravity \cite{Bergshoeff:2009hq}, pure fourth-order gravity \cite{Deser:2009hb} or generalizations thereof. 

\paragraph{Concluding remarks.}
We conclude with some remarks concerning four dimensions.
The Gross--Perry--Yaffe instability of 4-dimensional HFS due to nucleation of Schwarzschild black holes \cite{Gross:1982cv} is qualitatively different from the instability discussed in the present work, since the former only involves static spacetimes. Our transition from HFS into FSC is also different from the well-known quantum creation of universes \cite{Zeldovich:1984vk,Rubakov:1984bh,Vilenkin:1984wp}, since the latter requires the presence of some form of matter, like a scalar field with non-vanishing self-interaction potential. 
Given these differences to previous constructions, it would be interesting to generalize our results to four (or higher) dimensions. This could be feasible, since also 4- (or higher-) dimensional AdS allows the construction of BTZ-like quotients, see \cite{Aminneborg:2008sa} for a careful analysis and references therein for the original literature. 

\acknowledgments

%We are grateful to Hamid Afshar, Alejandra Castro, Reza Fareghbal, Michael Gary, Gaston Giribet, Gim Seng Ng, Jan Rosseel, Friedrich Sch\"oller and Ricardo Troncoso for discussions.
%
We are grateful to Hamid Afshar, Alejandra Castro, Michael Gary, Gaston Giribet, Gim Seng Ng, Jan Rosseel, Friedrich Sch\"oller, Ricardo Troncoso for discussions, and especially to Reza Fareghbal for collaboration during the initial stages of this work.
DG dedicates this paper to the memory of Gerhard Adam and numerous discussions about thermodynamics with him.
AB thanks the Center for Theoretical Physics, MIT for hosting him during this work. AB's stay at MIT was sponsored by an Early Career grant awarded jointly by the Scottish University Physics Alliance and the Edinburgh Partnership in Engineering and Mathematics. 
The work of JS and AB was partially supported by the Engineering and Physical Sciences Research Council (EPSRC) [grant number EP/G007985/1] and the Science and Technology Facilities Council (STFC) [grant number ST/J000329/1]. 
SD was supported by the Fundamental Laws Initiative of the Center for the Fundamental Laws of Nature, Harvard University. DG was supported by the START project Y435-N16 of the Austrian Science Fund (FWF) and the FWF project P21927-N16. 
AB and DG thank the Galileo Galilei Institute for Theoretical Physics for the hospitality and the INFN for partial support during the completion of this work.

\end{document}